\documentstyle[12pt, epsf]{article}
\newcommand\ZZZ{{\hbox{ Z\kern-1.6mm Z}}}
\newcommand{\beq}{\begin{equation}}
\newcommand{\eeq}{\end{equation}}
\newcommand{\bea}{\begin{eqnarray}}
\newcommand{\eea}{\end{eqnarray}}
\newcommand{\ra}{\rangle}
\newcommand{\la}{\langle}
\newcommand{\lt}{\left}
\newcommand{\rt}{\right}
\newcommand{\Iop}{\relax{\rm I\kern-.18em I}}
\newcommand{\one}{{\hbox{ 1\kern-1.2mm l}}}
\newcommand{\T}{{\cal T}}
\newcommand{\Tt}{{\tilde {\cal T}}}

\newcommand{\del}{\partial}

\newcommand{\s}{\sigma}
\newcommand{\eq}{\dot =}

\newcommand{\sectiono}[1]{\section{#1}\setcounter{equation}{0}}

\begin{document}
{}~
{}~
\hfill\vbox{\hbox{UK/08-06}}
\break

\vskip 2cm

\centerline{\Large \bf On the Conformal Field Theories for} 
\centerline{\Large \bf Bosonic Strings in PP-Waves}

\medskip

\vspace*{4.0ex}

\centerline{\large \rm Partha Mukhopadhyay }

\vspace*{4.0ex}

\centerline{\large \it Department of Physics and Astronomy}
\centerline{\large \it University of Kentucky}
\centerline{\large \it Lexington, KY 40506}

\medskip

\centerline{E-mail: partha@pa.uky.edu}

\vspace*{5.0ex}

\centerline{\bf Abstract} \bigskip

Recently Kazama and Yokoi (arXiv:0801.1561 [hep-th]) have used a phase-space method to study the Virasoro algebra of type IIB superstring theory in the maximally supersymmetric R-R plane wave background in a semi-light-cone gauge. Two types of normal ordering have been considered, namely ``phase space normal ordering" (PNO) and ``massless normal ordering" (MNO). The second one, which is the right one to choose in flat background, has been discarded with the argument that the Virasoro algebra closes only in the first case. To understand this issue better with a completely covariant treatment we consider the easiest case of bosonic strings propagating in an arbitrary pp-wave of the simplest kind. Using the phase-space method we show that MNO is in fact the right one to choose because of the following reason. For both types of normal ordering the energy-momentum tensor satisfies the desired Virasoro algebra up to anomalous terms proportional to the space-time equation of motion of the background. However, it is MNO which gives rise to the correct spectrum - we compute the quadratic space-time action by restricting the string field inside a transverse Hilbert space. This turns out to be non-diagonal. Diagonalizing this action reproduces the spectrum directly obtained in light-cone quantization.
The same method with PNO gives rise to a spectrum with negative dimensions.

\newpage

\tableofcontents

\baselineskip=18pt

\sectiono{Introduction and summary}
\label{s:intro}

Pp-waves are certain special gravitational backgrounds for which string theory can be more tractable \cite{stringpp, tseytlin}.
It is interesting to understand the conformal field theories (CFT) relevant to such backgrounds in general. Because the world-sheet theory is interacting the standard canonical quantization is not straightforward. It has recently been pursued in \cite{chizaki} for certain NS-NS pp-waves. More recently Kazama and Yokoi (KY) \cite{kazama} have applied a phase-space method to study type IIB superstring theory in the maximally supersymmetric R-R plane wave background \cite{RRpp}. The basic method involves imposing the equal time canonical commutation relations (after choosing the covariant gauge) to all the phase-space variables. This algebraic structure is independent of the dynamics of the theory and can be used to define a Hilbert space isomorphic to that corresponding to the flat background\footnote{In the usual canonical quantization method, where the Euler-Lagrange equations need to be solved, such an algebraic structure emerges after solving the theory.}. The dynamical questions can then be answered by writing the composite operators in terms of the phase-space variables and using the Heisenberg equation of motion. There are certain advantages of this method, at least for the present context, e.g. using this algebraic structure one can verify Virasoro algebra and establish conformal invariance \cite{kazama}.

There is certainly an issue of normal ordering which goes to the definition of the theory. Two types of normal ordering were considered in \cite{kazama} which are called {\it phase-space normal ordering} (PNO) and {\it massless normal ordering} (MNO). It turns out that MNO defines the theory correctly in flat background. However, it was found in \cite{kazama} that PNO is the right choice for the R-R plane wave as with MNO Virasoro algebra picks up anomalous terms. It was concluded in \cite{kazama}
that the theory does not have a smooth flat space limit. It is, however, not very clear why such a limit should not be smooth. Moreover, according to \cite{universality} there exists a universal sector of operators whose vacuum expectation values should turn out to be same as that in flat background. This will certainly not be true if the theories in these two backgrounds are defined with different types of normal ordering.

The analysis in \cite{kazama} has been done in a semi-light-cone gauge where bosons are treated covariantly, but the fermions are not. To avoid this complication and to understand the phase-space method better we consider bosonic string theory in the simplest class of pp-waves where no fluxes are switched on\footnote{More precisely, these pp-waves are given in eq.(\ref{simpmetric}). Unless stated otherwise, we will always refer to these particular backgrounds as ``pp-waves'' in the rest of this paper.}. We argue that with both MNO and PNO the Virasoro algebra is satisfied up to anomalous terms that are proportional to the equation of motion for the background\footnote{We present the details of the computation only for the case of MNO.}. With this result and the above argument we conclude that MNO defines the world-sheet theory correctly also for the pp-waves. It would be interesting to generalize this result to bosonic strings in arbitrary pp-waves which allow more general metric and include fluxes. In the context of type IIB R-R plane wave it will be interesting to see if a supersymmetric generalization of this result can be found maintaining the full covariance.

To establish further that MNO is the right choice we proceed to compute the physical spectrum of the theory by specializing to cases where the corresponding light-cone gauge fixed world-sheet theory is free. To obtain this spectrum using our CFT formulation we consider computing the closed string field theory (CSFT) \cite{csft, sen, zwiebach} kinetic term $S_2$ by restricting the string field to a transverse Hilbert space relevant to the light-cone gauge. Unlike in flat case the result turns out to be non-diagonal and the spectrum is computed by diagonalizing the action. This matches precisely with the spectrum obtained directly in light-cone gauge up to terms proportional to the equation of motion for the background. The same computation gives rise to a spectrum with negative dimensions if we define the theory with PNO.

The paper is organized as follows: We analyze the Virasoro algebra for the class of pp-waves of our interest and compute the physical spectrum for the suitable cases in section \ref{s:simplest}. All these results are obtained by using MNO. We explain the problems with PNO in section \ref{s:problems}. Some technical details are given in the appendices.

\section{Simplest pp-waves with MNO}
\label{s:simplest}

We consider closed bosonic string theory in a class of arbitrary background metric. Unless explicitly specified, we will follow the notations and conventions of \cite{polchinski}. When the background is an exact string solution we expect the matter part of the world-sheet theory to be a CFT such that the left and right moving Virasoro generators $L_n$ and $\tilde L_n$ respectively, which mutually commute, satisfy the following Virasoro algebra with central charge $c=26$.
\bea
[L_{m}, L_{n}] &=& (m-n) L_{m+n} + {c\over 12}
(m^{3}-m) \delta_{m+n,0}~.
\label{vir-st}
\eea

The class of backgrounds we wish to consider are the simplest pp-waves in $(d+2)$-dimensions:
\bea
ds^{2} = 2dx^{+}dx^{-} + K(x^+,\vec x) (dx^{+})^{2} + d\vec x. d\vec x~,
\label{simpmetric}
\eea
where $\vec x= (x^{1}, x^{2}, \cdots , x^{d})$ and $x^{\pm} = {1\over \sqrt{2}} (x^{d+1}\pm x^{0})$.
It is known \cite{stringpp} that (\ref{simpmetric}) is an exact string background provided:
\bea
d=24~, \hbox{ and } {\vec \del}^{2} K(x^{+},\vec x ) = 0 ~.
\label{cond}
\eea
Therefore the matter part of the world-sheet theory should realize the Virasoro algebra as described above whenever the condition (\ref{cond}) is satisfied. This is what we will demonstrate using the phase-space formulation of \cite{kazama}. We have given the details of this method for the flat background in appendix \ref{a:flat}. We follow the same steps in the present case.
The components of the conjugate momentum $P_{\mu}$ are now given by:
\bea
P_{-}= T \del_{\tau}X^{+} ~, ~~P_{+}= T \lt( \del_{\tau}X^{-} +K(X^{+},\vec X) \del_{\tau} X^+\rt)~,
~~P_{I} = T \del_{\tau}X_{I}~.
\label{P}
\eea
Defining $\Pi^{\pm},~ \Pi^I$ and the corresponding right moving counterparts in the same way as in eqs.(\ref{Pi-PX}) one finds that the left and
right moving components of the energy-momentum (EM) tensor take the following form in the present case:
\bea
\T = \T^{(0)} + \Delta \T ~, \quad \Tt  = \Tt^{(0)}+ \Delta \T ~,
\label{TTtilde}
\eea
where $\T^{(0)}$ and $\tilde \T^{(0)}$ are the EM tensor components for the flat background as given in (\ref{T0Ttilde0}) (we continue to use the notation (\ref{A.B})) and
\bea
\Delta \T &=& -{1\over 2} K(X^{+},\vec X) \Pi^{+} \tilde \Pi^{+}~,
\label{DeltaT}
\eea

To quantize the system one imposes the same equal time commutation relations as in (\ref{XPcomm}). All the equations (\ref{XPcomm}) through (\ref{hilbert}) are also valid in the present context. The complete set of states in (\ref{hilbert}) at a fixed instant of time describes the full Hilbert space of the matter CFT.

\subsection{Virasoro algebra}
\label{ss:vir}

The Virasoro modes $L_n$ and $\tilde L_n$ for the pp-waves are defined by Fourier expanding $\T$ and $\Tt$ in (\ref{TTtilde}) in the similar fashion as in (\ref{T0mode-expand})\footnote{As explained at the beginning of subsection (\ref{sa:massless}), we will be suppressing the $\tau$ dependence from all the variables.}. Below we will use MNO, as defined in appendix \ref{a:flat}, to show that $L_n$ and $\tilde L_n$ satisfy the Virasoro algebra (\ref{vir-st}) with central charge $c=d+2$ whenever the metric (\ref{simpmetric}) satisfies the space-time equation of motion as given in (\ref{cond}).

Let us start with the left moving sector. We first notice that given the Fourier expansion for $\T^{(0)}(\s)$ as in eqs.(\ref{T0mode-expand}) and the Virasoro algebra (\ref{vir-st}) satisfied by $L^{(0)}_n$ with $c=d+2$, one can show that the following identity holds:
\bea
[\T^{(0)}(\s),\T^{(0)}(\s')] &=& \pi i \lt[-{d+2\over 6} \delta'''(\s-\s') + \lt( 4 \T^{(0)}(\s)- {d+2\over 6} \rt) \delta'(\s-\s') \rt. \cr
&& \lt. + 2\del_{\s}\T^{(0)}(\s)\delta(\s-\s')\rt]~.
\label{T0comm}
\eea
We will use this observation in the reverse order to establish that $L_n$ satisfies the same Virasoro algebra (\ref{vir-st}) under the required condition. This requires us to argue that $\T(\s)=\T^{(0)}(\s)+\Delta \T(\s)$ also satisfies the same commutation relation in (\ref{T0comm}) with $\T^{(0)}(\s)$ replaced by $\T(\s)$. It has been shown in appendix \ref{a:comm} that,
\bea
[\Delta \T(\s), \Delta \T(\s')] = 0~.
\label{DeltaTcomm}
\eea
We therefore need to establish,
\bea
[\T^{(0)}(\s), \Delta \T(\s')] + [\Delta \T(\s), \T^{(0)}(\s')]
&=& 4 \pi i  \Delta \T(\s) \delta'(\s-\s') + 2\pi i \del_{\s} \Delta \T(\s) \delta(\s-\s')~, \cr &&
\label{T0DeltaTcomm}
\eea
whenever,
\bea
\vec \del^2 K(\s)  &\equiv & \vec \del^2 K(X^+(\s),\vec X(\s)) \cr
&=& \int {dk^- d \vec k\over (2\pi)^{(d+1)/2}} ~\tilde K(k^-,\vec k) ~\vec k^2
:e^{i(k^- X^+ +\vec k. \vec X)}(\s): \cr
&=& 0~,
\label{norm-ord-cond}
\eea
where $\tilde K(k^-,\vec k)$ is the Fourier transform of $K(x^+,\vec x)$ and $:\,\,:$ is the massless normal ordering. This is indeed true as shown in appendix \ref{a:comm}.

The analysis for the right moving sector proceeds in the similar way. In this case $\tilde {\cal T}^{(0)}(\s)$ satisfies:
\bea
[\tilde \T^{(0)}(\s),\tilde \T^{(0)}(\s')] &=& \pi i \lt[{d+2\over 6} \delta'''(\s-\s') - \lt( 4 \T^{(0)}(\s) - {d+2\over 6} \rt) \delta'(\s-\s') \rt. \cr
&& \lt. - 2\del_{\s}\T^{(0)}(\s)\delta(\s-\s')\rt]~.
\label{Ttilde0comm}
\eea
Therefore we need to show:
\bea
[\tilde \T^{(0)}(\s), \Delta \T(\s')] + [\Delta \T(\s), \tilde \T^{(0)}(\s')]
&=& -4 \pi i  \Delta \T(\s) \delta'(\s-\s') - 2\pi i \del_{\s} \Delta \T(\s) \delta(\s-\s')~, \cr &&
\label{Ttilde0DeltaTcomm}
\eea
whenever (\ref{norm-ord-cond}) is satisfied. This can be proved following the same procedure as in \ref{sa:proofT0DeltaTcomm}.

Finally, we need to show that the left and right moving Virasoro generators commute with each other. We have shown in appendix \ref{a:comm} that indeed,
\bea
[\T(\s), \Tt(\s')] = 0 ~,
\label{TTtildecomm}
\eea
whenever the condition (\ref{norm-ord-cond}) is satisfied.

\subsection{Physical spectrum}
\label{ss:physical}

In this section we will use our CFT method to reproduce the physical spectrum of strings obtained in light-cone gauge by specializing to the following cases:
\bea
K(x^+,\vec x) = - \sum_I s_I x^I x^I~, \hbox{ such that } \sum_I s_I =0~.
\label{special}
\eea
The second equation is the space-time equation of motion in (\ref{cond}) for the present ansatz.
The light-cone quantization can be done following \cite{polchinski}. In this case the transverse coordinate $X^I$ becomes a massive scalar field with mass squared \cite{RRpp}:
\bea
m_I^2 = s_I (\alpha' p^+)^2~,
\eea
$p^+$ being the light-cone momentum\footnote{Certainly, due to the condition in (\ref{special}), it is necessary that
some of the directions have negative mass squared. This implies that the system is unstable. This is a property of the background, chosen for simplicity, and can be bypassed by switching on fluxes. However, all the non-zero modes $X^I_n$'s (\ref{fields-mode-exp}), which we will be mainly concerned with, can be stabilized by keeping $|m_I^2|\leq 1$. In any case, here we are focussing on a method of analyzing string theory in pp-waves for which we may ignore such issues.}.
The light-cone hamiltonian $H_{lc}$ turns out to satisfy:
\bea
2 p^+ H_{lc} + \vec p^2 + \sum_I s_I X^I_0 X^I_0 (p^+)^2 + {2\over \alpha'} \lt[ \sum_{n>0,I} \lt(
\alpha^I_{-n} \alpha^I_n + \tilde \alpha^I_{-n} \tilde \alpha^I_n \rt) + {\cal C}(m_I^2)\rt] = 0~,
\label{Hlc}
\eea
where the non-trivial commutators of the oscillators are given by,
\bea
[\alpha^I_n, \alpha^J_m] = [\tilde \alpha^I_n, \tilde \alpha^J_m] = w^I_n \delta^{IJ} \delta_{n, -m}~, \quad
w^I_n = n\sqrt{1+ {m_I^2\over n^2}}~,
\label{alphacomm}
\eea
and the ``zero point energy" ${\cal C}(m_I^2)$ is given by the total Casimir energy of all the massive scalar fields:
\bea
{\cal C}(m_I^2) = \sum_{n>0, I} w^I_n~.
\label{casimir}
\eea
The expression (\ref{Hlc}) reduces to the correct one for flat background in the limit $s_I \to 0$.
In this case the zero point energy reduces to \cite{polchinski}:
\bea
\lim_{s_I\to 0} {\cal C}(m_I^2) \to d \sum_{n>0} n = -2 ~,
\label{zeta}
\eea
where use has been made of the Riemann zeta function: $\zeta (-1)= \sum_{n>0} n = -{1\over 12}$.

We will now derive an expression for the light-cone spectrum using our CFT approach and see that the result is indeed the same up to a term proportional to the space-time equation of motion in (\ref{special}). To do that we consider computing the CSFT kinetic term \cite{csft, sen, zwiebach}. Given any matter CFT of central charge $c=26$ this is given by \cite{sen},
\bea
S_2 = -{1\over 2} \la \Psi_{bpz} | c_0^- Q_B |\Psi \ra~,
\label{S2}
\eea
where $c^-_0=(c_0- \tilde c_0)/2$ and $Q_B$ is the total BRST operator\footnote{$Q_B$ is given by the same oscillator expression as in \cite{polchinski} with the matter Virasoro generators replaced by the ones defined in subsection \ref{ss:vir} for pp-waves.}.
$|\Psi \ra$ is the string field given by an arbitrary ghost number 2 state in the combined matter-ghost CFT satisfying:
\bea
(b_0-\tilde b_0)|\Psi\ra = (L^T_0-\tilde L^T_0)|\Psi\ra =0~,
\label{b0L0cond}
\eea
where $L^T_0$ and $\tilde L^T_0$ are the total (matter + ghost) Virasoro zero modes. $\la \Psi_{bpz}|$ is the BPZ conjugate of $|\Psi\ra$ as defined in \cite{zwiebach}.

Given the above definition of $S_2$ we will now discuss how to compute it using the phase-space method in the present context. Let us first consider the string field $|\Psi\ra$. Using eqs.(\ref{TTtilde}) one can see that the subsidiary conditions in (\ref{b0L0cond}) are same as that in flat background. Given this and the isomorphism between the CFT Hilbert spaces corresponding to flat and pp-wave backgrounds we conclude that $|\Psi\ra$ can be expanded in terms of the CFT states precisely in the same way as in flat case. Next we consider the computation of $\la \Psi_{bpz}|$. Usually BPZ conjugate of a state is defined through a conformal transformation given by the inversion map\footnote{On the euclidean world-sheet parameterized by  $(\tau_E,\s)$, $\tau_E$ being the euclidean time, the inversion map is given by $I\circ z = {1\over z}$, where $z=\exp(\tau_E-i\s)$.} $I$ applied on the relevant vertex operator\cite{zwiebach}. In our case it is convenient to view $I$ as the combined $PT$ transformation on the world-sheet with Minkowski signature:
\bea
I\circ \tau = -\tau~, \quad \quad I \circ \s = -\s~.
\eea
Its action on the phase-space modes can be easily determined to be\footnote{The action of $I$ on the ghost sector is taken to be the same as given in \cite{zwiebach}.}:
\bea
I\circ \Pi^{\mu}_n(\tau) = -\Pi^{\mu}_{-n}(\tau)~, \quad \quad I\circ \tilde \Pi^{\mu}_n(\tau) = -\tilde \Pi^{\mu}_{-n}(\tau)~,
\label{IonPi}
\eea
where we have explicitly displayed the $\tau$ dependence. Given any state $|\phi, k \ra = \phi |k\ra$, where $\phi$ contains all the matter and ghost creation operators, the BPZ conjugate is given by: $\la k, \phi_{bpz}|=\la -k| I\circ \phi$, where $\la k|$ is the hermitian conjugate of $|k\ra$, implying: $\la k|p\ra \propto \delta^{d+2}(k-p)$. Noticing that eqs.(\ref{IonPi}) are valid both in flat background and pp-waves we arrive at the following conclusion: In flat background the following relation holds \cite{zwiebach}:
\bea
\la \Psi_{bpz} | = - \la \Psi |~,
\label{reality}
\eea
$\la \Psi|$ being the hermitian conjugate of $|\Psi \ra$, whenever the string field components satisfy the correct reality property. This relation is still valid in pp-waves if the string field components satisfy the same reality condition. This enables us to write,
\bea
S_2 = {1\over 2} \la \Psi | c_0^- Q |\Psi \ra~,
\label{S2-2}
\eea
also for the pp-waves\footnote{Notice that in the phase-space formulation the matter part (\ref{hilbert}) of a CFT state is $\tau$ dependent. But $S_2$ itself is $\tau$ independent as expected. The reason is simply the fact that the matter part of any term in $S_2$ can be written as the vacuum expectation value of a collection of phase-space modes at the same time. Such an expectation value can be computed entirely by using the equal-time commutators in (\ref{fields-mode-comm}).}.

For the purpose of computing the light-cone spectrum we will now compute $S_2$ following eq.(\ref{S2-2}) by restricting the string field to the relevant transverse Hilbert space ${\cal H}_{\perp}$\footnote{Since the standard light-cone gauge can be fixed on the world-sheet, it is expected that all the BRST cohomology elements can be found in this transverse space. See
also \cite{kazama}.}. ${\cal H}_{\perp}$ is spanned by
the following basis states which satisfy the $(b_0-\tilde b_0)$-condition in (\ref{b0L0cond}),
\bea
|\{N_{In}\}, \{\tilde N_{In}\}, p ) &\equiv & c_1 \tilde c_1 |\{N_{In}\}, \{\tilde N_{In}\}, p \ra ~, \cr && \cr
|\{N_{In}\}, \{\tilde N_{In}\}, p \ra & \equiv & \prod_{n>0,I} {(\Pi^I_{-n})^{N_{In}} (\tilde \Pi^I_{-n})^{\tilde N_{In}}
\over \sqrt{ n^{N_{In}+\tilde N_{In}} N_{In}! \tilde N_{In}!}} |p\ra~,
\label{basis}
\eea
The restricted string field is therefore expanded as:
\bea
|\Psi_{\perp}\ra = \sum'_{\{N_{In}\}, \{\tilde N_{In}\}} \int dk~
\tilde \psi_{\{N_{In}\}, \{\tilde N_{In}\}}(k)  ~|\{N_{In}\}, \{\tilde N_{In}\}, p )~,
\label{Psiperp}
\eea
where the prime on the summation refers to the restriction $\sum_{n>0,I}n(N_{In}-\tilde N_{In}) =0 $ coming from the $(L^T_0-\tilde L^T_0)$-condition in (\ref{b0L0cond}). Following the standard computation the quadratic action is evaluated to be,
\bea
S_2 &\propto& \sum'_{\{N'_{In}\}, \{\tilde N'_{In}\}} \sum'_{\{N_{In}\}, \{\tilde N_{In}\}} \int dp' dp~
\tilde \psi^*_{\{N'_{In}\}, \{\tilde N'_{In}\}}(p') \tilde \psi_{\{N_{In}\}, \{\tilde N_{In}\}}(p)\cr
&& \la p', \{N'_{In}\}, \{\tilde N'_{In}\}|  {\cal S}_2 |\{N_{In}\}, \{\tilde N_{In}\},p\ra ~,
\label{S2nondiag}
\eea
where $\tilde \psi^*_{\{N_{In}\}, \{\tilde N_{In}\}}(p)$ is the complex conjugate of $\tilde \psi_{\{N_{In}\}, \{\tilde N_{In}\}}(p)$, $\la p, \{N_{In}\}, \{\tilde N_{In}\}|$ is the hermitian conjugate of $|\{N_{In}\}, \{\tilde N_{In}\}, p\ra$ and
\bea
{\cal S}_2 &=& L_0+ \tilde L_0 -2 = {\cal S}^0_2 + {\cal S}^{\neq 0}_2~, \cr
{\cal S}^0_2 &=& 2 p^+ p^- + \vec p^2 + \sum_I s_I X^I_0 X^I_0 (p^+)^2 ~, \cr
{\cal S}^{\neq 0}_2 &=&  {2\over \alpha'} \sum_{n>0,I} \lt\{
(1+{m_I^2\over 2 n^2}) (\Pi^I_{-n} \Pi^I_n +\tilde \Pi^I_{-n} \tilde \Pi^I_n) - {m_I^2\over 2 n^2} (\Pi^I_n \tilde \Pi^I_n
+ \Pi^I_{-n}\tilde \Pi^I_{-n}) \rt\} - {4\over \alpha'}  ~, \cr &&
\label{calS2}
\eea
where the second and third lines display the contribution of $(L_0+\tilde L_0-2)$ that is nonzero inside ${\cal H}_{\perp}$.
Obviously the quadratic action is not diagonal in the basis that we are working with. To diagonalize this we
relate the $\Pi$-oscillators and the $\alpha$-oscillators in (\ref{alphacomm}) in the following way:
\bea
\Pi^I_n={1\over 2} (M^+_{In}\alpha^I_n + M^-_{In} \tilde \alpha^I_{-n})~, \quad
\tilde \Pi^I_n={1\over 2} (M^+_{In} \tilde \alpha^I_n + M^-_{In} \alpha^I_{-n})~, \quad \forall n\neq 0~,
\label{Pi-alpha}
\eea
where, $M^{\pm}_{In} = 1 \pm {n\over w^I_n}$.
One can check that ${\cal S}^{\neq 0}_2$ is diagonal in terms of the $\alpha$-oscillators and it takes the following form
\bea
{\cal S}^{\neq 0}_2 &=& {2\over \alpha'} \lt[ \sum_{n>0,I} \lt(\alpha^I_{-n} \alpha^I_n + \tilde \alpha^I_{-n} \tilde \alpha^I_n \rt)
+\hat{\cal C}(m_I^2)\rt] ~,
\label{calS2diag}
\eea
The light-cone hamiltonian $H_{lc}$ is identified with $p^-$ in the second equation of (\ref{calS2}) and therefore is given by the same expression (\ref{Hlc}) with ${\cal C}(m_I^2)$ replaced by $\hat{\cal C}(m_I^2)$ which is found to be,
\bea
\hat{\cal C}(m_I^2) &=& -2 + {1\over 2} \sum_{n>0,I} w^I_n \lt[\lt(1+{m_I^2\over 2n^2}\rt)(M^-_{In})^2 - {m_I^2\over 2n^2} M^+_{In} M^-_{In} \rt]~, \cr
&=& -2 - d \sum_{n>0} n + {\cal C}(m_I^2) - {1\over 2}\lt( \sum_I m_I^2 \rt) \sum_{n>0} {1\over n}~,
\label{calChat}
\eea
The first two terms in the second line cancel according to (\ref{zeta}). The last term drops off when the space-tme equation of motion
in (\ref{special}) is satisfied, in which case $\hat {\cal C}(m_I^2)= {\cal C}(m_I^2)$.

Certainly the quadratic action takes the diagonal form in terms of a new set of fields  $\tilde \chi_{\{N_{In}\}, \{\tilde N_{In}\}}(p)$ which correspond to the basis states $|\{N_{In}\},\{\tilde N_{In}\}, k)) \equiv c_1 \tilde c_1 |\{N_{In}\},\{\tilde N_{In}\}, k\ra \ra$, where $|\{N_{In}\},\{\tilde N_{In}\}, k\ra \ra $ is given by the second equation in (\ref{basis}) with all the $\Pi$ oscillators replaced by the corresponding $\alpha$ oscillators and
$|k\ra$ replaced by the new vacuum $|k\ra\ra$ defined by,
\bea
\alpha^I_n |k\ra \ra = \tilde \alpha^I_n |k\ra \ra = 0~, \quad \forall n>0~,
\label{new-vac}
\eea
Given this definition and eqs.(\ref{Pi-alpha}) it is clear that $|k\ra$ is given by a squeezed state in the Fock space over $|k\ra\ra$ (and {\it vice versa}) with a normalization constant that can be fixed by demanding: $\la k'| k \ra = \la\la k'|k\ra\ra$.
Both the basis $|\{N_{In}\},\{\tilde N_{In}\}, k\ra$ and $|\{N_{In}\},\{\tilde N_{In}\}, k\ra \ra$ are orthonormal and complete. Using this one can explicitly relate the two sets of fields $\tilde \psi_{\{N_{In}\}, \{\tilde N_{In}\}}(k)$ and $\tilde \chi_{\{N_{In}\}, \{\tilde N_{In}\}}(k)$. However we will not need those expressions for the purpose of the present work.

\section{Problem with PNO}
\label{s:problems}

The phase-space normal ordering \cite{kazama} is given by identifying $X^{\mu}_n$'s with $n>0$ and $P_{\mu n}$'s with $n\geq 0$ as annihilation operators. Using the relation (\ref{Pi-PX}) and the Fourier expansions (\ref{fields-mode-exp}) we conclude that the corresponding ground state $|0\ra'$ satisfies:
\bea
\Pi^{\mu}_n |0\ra' = \tilde \Pi^{\mu}_{-n} |0\ra' =0~, \quad \forall n\geq 0~.
\label{vac-prime}
\eea
Therefore $|0\ra'$ differs from $|0\ra$ in the right moving sector. By redefining the composite operators with PNO and repeating the similar analysis of subsection \ref{ss:vir} and appendix \ref{a:comm} one can show that the EM tensor components satisfy the required Virasoro algebra up to anomalous terms proportional to the equation of motion for the background.
We will not discuss this in any further detail. Instead we will show in this section that such a normal ordering gives rise to an incorrect spectrum.

To see this let us first consider the flat background. Since $L^{(0)}_n$ and $\tilde L^{(0)}_n$ do not involve any normal ordering for $n\neq 0$, they are still given by the same expressions as in the case of MNO. Therefore we have,
\bea
L^{(0)}_n|0\ra'= \tilde L^{(0)}_{-n} |0\ra' =0 ~, \quad \forall n \geq 0~.
\eea
Certainly this is not the correct choice of the vacuum. In particular, the excited states (right moving sector) are obtained by applying $\tilde L^{(0)}_n$'s ($n>0$) on $|0\ra'$ and have negative conformal dimensions. Notice that $|0\ra'$ can not be constructed in the Fock space built over $|0\ra$ and {\it vice versa} as the latter contains states of positive dimensions only.

The fact that PNO is a wrong choice for the flat background already tells us that it does not work for the pp-wave background, as it does not lead to the right flat space limit. Another way to see this is the following: There exists a universal sector of operators \cite{universality} to which the background appears to be flat. This set of operators corresponds to a subspace of the full Hilbert space whose overlap with ${\cal H}_{\perp}$ is given by all the states in ${\cal H}_{\perp}$ with zero light-cone momentum $p^+=0$. It is a necessary condition that the matrix elements of the ``universal operators'' between the states inside this overlap should turn out to be same as that computed in flat background. This condition is certainly not satisfied if we define the theory with different normal ordering for flat background and pp-waves.

We will now show that if we compute $S_2$ following the same steps of subsection \ref{ss:physical} then, just like in flat space, we get a spectrum of negative dimensions. Now the transverse sector ${\cal H}_{\perp}$ has to be spanned by the basis states $|\{N_{In}\}, \{\tilde N_{In}\}, k )' = c_1 \tilde c_1 |\{N_{In}\}, \{\tilde N_{In}\}, k \ra'$, where the states $|\{N_{In}\}, \{\tilde N_{In}\}, k \ra'$ are obtained by replacing $\tilde \Pi^I_{-n} \to \tilde \Pi^I_n$ and $|k\ra \to |k\ra'= e^{ik.X_0}|0\ra'$ in the last equation of (\ref{basis})\footnote{The state $|\{N_{In}\}, \{\tilde N_{In}\}, k \ra'$, therefore, has a negative norm whenever $\sum_{n>0,I}\tilde N_{In}$ is odd.}. The result for $S_2$ is given by eq.(\ref{S2nondiag}) with the old basis replaced by the new ones. For ${\cal S}_2$, the first two equations in (\ref{calS2}) still hold. But the non-zero-mode-contribution ${\cal S}^{\neq 0}_2$ is given by the same expression in (\ref{calS2}) with opposite ordering for the $\tilde \Pi$ oscillators, which is expected, given eqs.(\ref{vac-prime}).
Diagonalizing such an expression one gets:
\bea
{\cal S}^{\neq 0}_2 &=& {2\over \alpha'} \sum_{n>0,I} \lt(\alpha^I_{-n} \alpha^I_n + \tilde \alpha^I_{n} \tilde \alpha^I_{-n} \rt)
-{4\over \alpha'} ~,
\label{S2PNOdiag}
\eea
Notice again that the $\tilde \alpha$-oscillators are oppositely ordered with respect to the $\alpha$-oscillators. This is because given
eqs. (\ref{vac-prime}) and (\ref{Pi-alpha}) one concludes,
\bea
\alpha^I_n |k\ra' = \tilde \alpha^I_{-n} |k\ra' = 0 ~, \quad \forall n>0~.
\eea
Given the commutators in (\ref{alphacomm}), it is clear that $S_2^{\neq 0}$ in (\ref{S2PNOdiag}) has negative eigenvalues\footnote{In fact both $L_0$ and $\tilde L_0$ separately have negative eigenvalues.}. We emphasize that the operators in (\ref{S2PNOdiag}) can not be normal ordered with respect to the $\alpha$-vacuum $|k\ra\ra$ defined in (\ref{new-vac}), as was done in \cite{kazama}. This is simply because the states $|k\ra'$ and $|k\ra \ra$ can not be constructed in the Fock space over each other.

\medskip
\centerline{\bf Acknowledgement}
\noindent

I thank S. Das for discussion.

\appendix

\section{Phase-space formulation for flat background}
\label{a:flat}

To introduce the phase-space formulation of \cite{kazama} and to set up our notations we work out the details for the flat background here. We will simply draw a parallel when we study pp-waves in sec.\ref{s:simplest}. For certain quantities we use the superscript $(0)$ to refer to the flat background. The same notations without this superscript will refer to the pp-waves. All the equations in this appendix where the two sides are related by $\eq$ will also be valid for pp-waves with the superscript $(0)$ removed wherever needed.

The world-sheet action for the flat background is: $S^{(0)} = - {T\over 2}\int d\tau d\sigma \sqrt{-g} g^{ab} \eta_{\mu \nu} \del_{a}X^{\mu} \del_{b} X^{\nu}$,
where $T={1\over 2\pi \alpha'}$ is the string tension and $\mu,\nu = 0,1,\cdots, d+1$. The energy-momentum (EM) tensor is defined to be: $\T^{(0)}_{ab}(\tau,\sigma) ~\eq - {4\pi \over \sqrt{-g}} {\delta S^{(0)} \over \delta g^{ab}(\tau, \sigma)}$.
After fixing the covariant gauge: $g_{ab}=\eta_{ab}$ the left and the right moving components of the EM tensor take the following form:
\bea
\T^{(0)} &\eq & {1\over 2} (\T_{\tau \tau} + \T_{\tau \sigma}) =  {1\over 2} \Pi . \Pi ~, \cr
\Tt^{(0)} &\eq & {1\over 2} (\Tt_{\tau \tau} - \Tt_{\tau \sigma}) = {1\over 2} \tilde \Pi . \tilde \Pi~,
\label{T0Ttilde0}
\eea
where we have used,
\bea
A.B ~\eq ~ A^+B^-+A^-B^++\vec A.\vec B~,
\label{A.B}
\eea
with the $\pm$ components defined below eq.(\ref{simpmetric}) and
\bea
\Pi^{\mu} ~\eq~  \sqrt{\pi \over T } \eta^{\mu \nu} P_{\nu} + \sqrt{\pi T} \del_{\sigma} X^{\mu} ~, \quad \tilde \Pi^{\mu} ~\eq~  \sqrt{\pi \over T } \eta^{\mu \nu} P_{\nu} - \sqrt{\pi T} \del_{\sigma} X^{\mu} ~,
\label{Pi-PX}
\eea
where
$P_{\mu} ~\eq~ {\delta{S^{(0)}}\over \delta{\del_{\tau}X^{\mu}}}$ is the momentum conjugate to $X^{\mu}$.
Notice that $\Pi^{\mu}$ and $\tilde \Pi^{\mu}$ are defined using $\eta^{\mu \nu}$ also for the pp-waves.

\subsection{Quantization with MNO}
\label{sa:massless}

In the phase-space method all the world-sheet fields and their conjugate momenta are Fourier expanded along the $\sigma$ direction at a given instant of time $\tau$. The complete operator algebra is also constructed at a given instant of time. This is sufficient to construct the Hilbert space. To obtain the time evolution of any operator one uses the Heisenberg equation of motion. Now onwards we will suppress the $\tau$ dependence of all the fields and it is understood that all the following analysis are done at a fixed instant of time. The equal time canonical commutation relations are:
\bea
[X^{\mu}(\sigma), P_{\nu}(\sigma')] ~\eq~ i \delta^{\mu}_{\nu} \delta(\sigma -\sigma')~.
\label{XPcomm}
\eea
The Fourier expansion of various fields are given by,
\bea
X^{\mu}(\sigma) ~\eq~ \sum_{n\in Z} X^{\mu}_n e^{-in \sigma}~, &\quad& P^{\mu}(\s) ~\eq~ \sum_{n\in Z} P^{\mu}_n e^{-in\s}~, \cr
\Pi^{\mu} (\s) ~\eq~ \sum_{n\in Z} \Pi^{\mu}_n e^{-in \s}~, &\quad& \tilde \Pi^{\mu}(\s) ~\eq~ \sum_{n\in Z} \tilde \Pi^{\mu}_n e^{in \s} ~,
\label{fields-mode-exp}
\eea
implying the hermiticity property: $(A^{\mu}_n)^{\dagger} ~\eq~ A^{\mu}_{-n}$, where $A^{\mu}_n$ stands for any of the Fourier modes appearing in the above equations and the following non-trivial commutators:
\bea
[X^{\mu}_m, P_{\nu n}]~\eq~ {i\over 2\pi} \delta^{\mu}_{\nu} \delta_{m+n,0}~,\quad
[\Pi^{\mu}_m, \Pi^{\nu}_n] ~\eq~ [\tilde \Pi^{\mu}_m, \tilde \Pi^{\nu}_n] ~\eq~ m \eta^{\mu \nu} \delta_{m+n,0}~.
\label{fields-mode-comm}
\eea
Therefore the complete set of operators are\footnote{For the non-zero modes the $\Pi$-oscillators defined here are same as the standard $\alpha$-oscillators (with the $\tau$-evolution included) conventionally used \cite{polchinski} for string quantization in flat background. The zero modes are related by a proportionality constant.}:
\bea
\hbox{zero modes: }&&  X^{\mu}_0~, \quad \eta^{\mu \nu} P_{\nu 0} ~\eq~\sqrt{T\over \pi} \Pi^{\mu}_0 ~\eq~ \sqrt{T\over \pi} \tilde \Pi^{\mu}_0~, \cr
\hbox{non-zero modes: } && \Pi^{\mu}_n~, \quad \tilde \Pi^{\mu}_n~, \quad \quad n \neq 0~.
\label{cso}
\eea
The massless normal ordering is defined by identifying $\Pi^{\mu}_n$ and $\tilde \Pi^{\mu}_n$ for $n\geq 0$ as annihilation operators. As it is well known that the corresponding ground state is indeed the SL(2,C) invariant vacuum $|0\ra$, i.e,
\bea
\Pi^{\mu}_n |0\ra  ~\eq~ \tilde \Pi^{\mu}_n |0\ra ~\eq~ 0 ~, \quad \forall n\geq 0~.
\label{massless-norm-ord}
\eea
The excited states are given by:
\bea
(\Pi^{\mu_1}_{-n_1} \Pi^{\mu_2}_{-n_2} \cdots)(\tilde \Pi^{\nu_1}_{-m_1} \tilde \Pi^{\nu_2}_{-m_2} \cdots) |k\ra~,
\label{hilbert}
\eea
where $|k\ra ~\eq~ e^{ik.X_0}|0\ra$. The Virasoro generators $L^{(0)}_n$ and $\tilde L^{(0)}_n$, which satisfy the Virasoro algebra (\ref{vir-st}) with central charge $c=d+2$, are defined by,
\bea
\T^{(0)}(\s) ~\eq~ \sum_{n\in Z} L^{(0)}_n e^{-in\s}~, \quad \tilde \T^{(0)}(\s) ~\eq~ \sum_{n\in Z} \tilde L^{(0)}_n e^{in\s}~.
\label{T0mode-expand}
\eea
The time evolution of various operators can be obtained by using the Heisenberg equation of motion:
\bea
{d\over d\tau} {\cal O} ~\eq~ i [{\cal O}, H^{(0)}] + {\del \over \del \tau} {\cal O}~,
\label{heisenberg}
\eea
where $H^{(0)} ~\eq~ 2\pi (L^{(0)}_0+\tilde L^{(0)}_0)$ is the hamiltonian. Using this one can show:
\bea
\Pi^{\mu}(\tau, \s) = \sum_{n\in Z} \Pi^{\mu}_n(0) e^{in(2\pi \tau -\s)}~,\quad \quad \tilde \Pi^{\mu}(\tau, \s) = \sum_{n\in Z} \tilde \Pi^{\mu}_n(0) e^{in(2\pi \tau +\s)}~,
\label{Pi-time}
\eea
\bea
\T^{(0)}(\tau, \s) ~\eq~ \sum_{n\in Z} L^{(0)}_n(0) e^{in(2\pi \tau -\s)}~,\quad \quad \Tt^{(0)}(\tau, \s) ~\eq~ \sum_{n\in Z} \tilde L^{(0)}_n(0) e^{in(2\pi \tau +\s)}~,
\label{T-time}
\eea
where we have shown the time dependence explicitly. Notice that the ``holomorphicity" properties of the fields
$\Pi^{\mu}$ and $\tilde \Pi^{\mu}$ in eqs.(\ref{Pi-time}) will not hold for pp-waves as the corresponding modes will have more complicated commutation relations with the hamiltonian. However, the same for $\T^{(0)}$ and $\Tt^{(0)}$ are obtained using the Virasoro algebra (\ref{vir-st}). Therefore eqs.(\ref{T-time}) (with the superscript $(0)$ removed) will be valid also in pp-waves.

\section{Relevant commutators}
\label{a:comm}

Below we prove equations (\ref{DeltaTcomm}, \ref{T0DeltaTcomm},  \ref{TTtildecomm}) required to establish the Virasoro algebra in subsection \ref{ss:vir}.
\subsection{Proof of Eq.(\ref{DeltaTcomm})}
\label{sa:proofDeltaTcomm}

The normal ordered expression for $\Delta \T(\s)$ is given by:
\bea
\Delta \T(\s) = -{1\over 2} K\Pi^+\tilde \Pi^+(\s)~,
\eea
where,
\bea
K(\s) &\equiv& :K(X^+(\s),\vec X(\s)): ~, \cr
&=& \int {dk^- d\vec k\over (2\pi)^{(d+1)/2}} ~\tilde K(k^-,\vec k) ~e^{ik^-X^+}(\s) :e^{i\vec k.\vec X}:(\s)~,
\eea
and
\bea
:e^{i\vec k.\vec X}:(\s) = e^{i\vec k.\vec X_0} E(\vec k;\s) \tilde E(\vec k; \s)~,
\eea
where,
\bea
E(\vec k;\s) &=& E_-(\vec k; \s) E_+(\vec k; \s)~,\cr
E_{\pm}(\vec k; \s) &=& \exp \lt(\mp {1 \over 2 \sqrt{\pi T}} \sum_{n\geq 1} {1\over n} \vec k. \vec \Pi_{\pm n} e^{\mp in\s} \rt)~.
\eea
The operator $ \tilde E(\vec k; \s)$ takes the similar form as above with $\Pi^I_{\pm n}$ replaced by $\tilde \Pi^I_{\pm n}$ and
$\s$ replaced by $-\s$. To evaluate the commutator $[\Delta \T(\s), \Delta \T(\s')]$ one needs to compute
$[:e^{i\vec k.\vec X}(\s):, :e^{i\vec p.\vec X}(\s'):]$. The result turns out to be zero as can be shown by using:
\bea
E(\vec k; \s) \tilde E(\vec k; \s) E(\vec p; \s')\tilde E(\vec p; \s') = :E(\vec k; \s) \tilde E(\vec k; \s) E(\vec p; \s')\tilde E(\vec p; \s') : e^{\vec k.\vec p D(\s-\s')}~,
\eea
where $D(\s)$ is an even function of $\s$.
This establishes eq.(\ref{DeltaTcomm}).

\subsection{Proof of Eq.(\ref{T0DeltaTcomm})}
\label{sa:proofT0DeltaTcomm}

Using the expressions in (\ref{TTtilde}), (\ref{T0Ttilde0}) and (\ref{DeltaT}) we may
write:
\bea
[\T^{(0)}(\s), \Delta \T(\s')] &=& -{1\over 2} [:\Pi^+\Pi^-(\s):, K\Pi^+\tilde \Pi^+(\s')]\cr
&& -{1\over 4} [:\vec \Pi. \vec \Pi(\s):, K\Pi^+\Pi^+(\s')]~.
\label{Ttilde0DeltaTcomm}
\eea
The normal ordered quadratic operators are given by,
\bea
:\Pi^{\mu}\Pi^{\nu}(\s): = \Pi^{\mu}_{(+)}\Pi^{\nu}_{(+)}(\s) + \Pi^{\mu}_{(-)}\Pi^{\nu}_{(-)}(\s) + \Pi^{\mu}_{(-)}\Pi^{\nu}_{(+)}(\s)
+ \Pi^{\nu}_{(-)}\Pi^{\mu}_{(+)}(\s)~,
\eea
where following \cite{kazama} we have defined,
\bea
\Pi^{\mu}_{(+)}(\s) = \sum_{n\geq 0} \Pi^{\mu}_n e^{-in\s}~, \quad
\Pi^{\mu}_{(-)}(\s) = \sum_{n> 0} \Pi^{\mu}_{-n} e^{in\s}~.
\eea
Given the above definitions the commutators in eq.(\ref{Ttilde0DeltaTcomm}) can be computed using eqs.(\ref{fields-mode-comm}) and the techniques used in \cite{kazama}.  The final result is found to be,
\bea
[\T^{(0)}(\s), \Delta \T(\s')] &=& \lt(-\pi i K(\s')\Pi^+(\s) \tilde
\Pi^+(\s')
+{i\over 8T} \vec \del^2 K \Pi^+\tilde \Pi^+(\s') \rt)
\delta'(\s-\s') \cr
&& + \sqrt{\pi\over T} \lt( {i\over 2} \del_+K (\Pi^+)^2 \tilde \Pi^+(\s)
+{i\over 2} :\vec \Pi.\vec \del K: \Pi^+ \tilde \Pi^+(\s) \rt. \cr
&& \lt. + {1\over 8 \sqrt{\pi T}} \vec \del^2 K \Pi^+\tilde \Pi^+(\s)
\rt) \delta(\s-\s')~,
\eea
where the derivatives of $K$ are defined following the same way as in (\ref{norm-ord-cond}).
Using this result and the identity \cite{kazama},
\bea
{\cal O}(\s') \delta'(\s-\s') = {\cal O} (\s) \delta'(\s-\s') +\del_{\s}
{\cal O}(\s) \delta (\s-\s')~,
\label{delta'}
\eea
one finds:
\bea
[\T^{(0)}(\s),\Delta \T(\s')] + [\Delta \T(\s), \T^{(0)}(\s')]
&=& 4\pi i \Delta \T(\s) \delta'(\s-\s') + 2\pi i \del_{\s} \Delta
\T(\s) \delta(\s-\s') \cr
&& -{i\over 4T}\lt(2 \vec \del^2 \Delta \T (\s) \delta'(\s-\s') \rt. \cr
&& \lt.+ \del_{\s} \vec \del^2 \Delta \T(\s) \delta(\s-\s') \rt) ~,
\eea
where
\bea
\vec \del^2 \Delta \T (\s) = -{1\over 2} \vec \del^2 K
\Pi^+ \tilde \Pi^+(\s)~.
\eea
This establishes eqs.(\ref{T0DeltaTcomm}, \ref{norm-ord-cond}).

\subsection{Proof of Eq.(\ref{TTtildecomm})}
\label{sa:proofTTtildecomm}

Using $[\T^{(0)}(\s), \tilde \T^{(0)}(\s')] = 0$ and (\ref{DeltaTcomm}) one can write,
\bea
[\T(\s),\tilde \T(\s')] = [\T^{(0)}(\s), \Delta \T(\s')] + [\Delta
\T (\s), \tilde T^{(0)}(\s')]~.
\eea
Computation of the commutators on the right hand side gives the following result:
\bea
[\T(\s),\tilde \T(\s')] &=& \pi i \lt( K\Pi^+(\s) \tilde \Pi^+(\s') +
{1\over 8 \pi T} \vec \del^2 K \Pi^+ \tilde \Pi^+(\s') \rt. \cr
&& \lt. -K\tilde \Pi^+(\s') \Pi^+(\s) -
{1\over 8 \pi T} \vec \del^2 K \Pi^+ \tilde \Pi^+(\s)\rt)
\delta'(\s-\s') \cr
&& +{i\over 2} \sqrt{\pi \over T} \lt(\del_+K (\Pi^+)^2 \tilde \Pi^+(\s)
+ :\vec \Pi.\vec \del K \Pi^+ \tilde \Pi^+(\s):  \rt. \cr
&&\lt. - \del_+K \Pi^+ (\tilde \Pi^+)^2(\s)
- :\vec{\tilde \Pi}.\vec \del K \Pi^+ \tilde \Pi^+(\s): \rt) \delta(\s-\s') ~.
\eea
Using the identity (\ref{delta'}) and $\Pi^{\mu}-\tilde \Pi^{\mu}= 2
\sqrt{\pi T} \del_{\s} X^{\mu}$ the above expression can be simplified to:
\bea
[\T(\s),\tilde \T(\s')] = {i\over 8T} \del_{\s} \lt(\vec \del^2 K \Pi^+ \tilde \Pi^+(\s)
\rt) \delta(\s-\s')~.
\eea
This establishes eqs.(\ref{TTtildecomm}, \ref{norm-ord-cond}).


\begin{thebibliography}{99}

\bibitem{stringpp}
D.~Amati and C.~Klimcik,
``NONPERTURBATIVE COMPUTATION OF THE WEYL ANOMALY FOR A CLASS OF NONTRIVIAL
BACKGROUNDS,''
Phys.\ Lett.\  B {\bf 219} (1989) 443;
G.~T.~Horowitz and A.~R.~Steif,
``SPACE-TIME SINGULARITIES IN STRING THEORY,''
Phys.\ Rev.\ Lett.\  {\bf 64}, 260 (1990);
R.~E.~Rudd,
``COMPACTIFICATION PROPAGATION,''
Nucl.\ Phys.\  B {\bf 352}, 489 (1991);
C.~Duval, Z.~Horvath and P.~A.~Horvathy,
``Vanishing of the conformal anomaly for strings in a gravitational wave,''
Phys.\ Lett.\  B {\bf 313}, 10 (1993)
[arXiv:hep-th/0306059];
C.~Duval, Z.~Horvath and P.~A.~Horvathy,
``Strings in plane-fronted gravitational waves,''
Mod.\ Phys.\ Lett.\  A {\bf 8}, 3749 (1993)
[arXiv:hep-th/0602128].

\bibitem{tseytlin}
A.~A.~Tseytlin,
``A Class of finite two-dimensional sigma models and string vacua,''
Phys.\ Lett.\  B {\bf 288}, 279 (1992)
[arXiv:hep-th/9205058];
A.~A.~Tseytlin,
``String Vacuum Backgrounds With Covariantly Constant Null Killing Vector And
2-D Quantum Gravity,''
Nucl.\ Phys.\  B {\bf 390}, 153 (1993)
[arXiv:hep-th/9209023];
A.~A.~Tseytlin,
``Finite Sigma Models And Exact String Solutions With Minkowski Signature
Metric,''
Phys.\ Rev.\  D {\bf 47}, 3421 (1993)
[arXiv:hep-th/9211061].

\bibitem{chizaki}
Y.~Chizaki and S.~Yahikozawa,
``General Operator Solutions and BRST Quantization of Superstrings in the
pp-Wave with Torsion,''
Prog.\ Theor.\ Phys.\  {\bf 118}, 1127 (2007)
[arXiv:0709.2991 [hep-th]];
Y.~Chizaki and S.~Yahikozawa,
``Covariant BRST Quantization of Closed Strings in the PP-Wave Background,''
Prog.\ Theor.\ Phys.\  {\bf 116}, 937 (2007)
[arXiv:hep-th/0608185].

\bibitem{kazama}
Y.~Kazama and N.~Yokoi, ``Superstring in the plane-wave background
with RR flux as a conformal field theory,'' JHEP {\bf 0803}, 057
(2008)
[arXiv:0801.1561 [hep-th]].

\bibitem{RRpp}
M.~Blau, J.~Figueroa-O'Farrill, C.~Hull and G.~Papadopoulos,
``A new maximally supersymmetric background of IIB superstring theory,''
JHEP {\bf 0201}, 047 (2002)
[arXiv:hep-th/0110242];
R.~R.~Metsaev,
``Type IIB Green-Schwarz superstring in plane wave Ramond-Ramond
background,''
Nucl.\ Phys.\  B {\bf 625}, 70 (2002)
[arXiv:hep-th/0112044];
R.~R.~Metsaev and A.~A.~Tseytlin,
``Exactly solvable model of superstring in plane wave Ramond-Ramond
background,''
Phys.\ Rev.\  D {\bf 65}, 126004 (2002)
[arXiv:hep-th/0202109].

\bibitem{universality}
P.~Mukhopadhyay,
``Tachyon condensation and non-BPS D-branes in a Ramond-Ramond plane wave
background,''
arXiv:hep-th/0611138;
P.~Mukhopadhyay,
``A Universality in PP-Waves,''
JHEP {\bf 0706}, 061 (2007)
[arXiv:0704.0085 [hep-th]].

\bibitem{csft}
M.~Saadi and B.~Zwiebach,
``CLOSED STRING FIELD THEORY FROM POLYHEDRA,''
Annals Phys.\  {\bf 192}, 213 (1989);
T.~Kugo, H.~Kunitomo and K.~Suehiro,
``Nonpolynomial Closed String Field Theory,''
Phys.\ Lett.\  B {\bf 226}, 48 (1989);
T.~Kugo and K.~Suehiro,
``Nonpolynomial Closed String Field Theory: Action And Its Gauge
Invariance,''
Nucl.\ Phys.\  B {\bf 337}, 434 (1990).

\bibitem{sen}
A.~Sen,
``ON THE BACKGROUND INDEPENDENCE OF STRING FIELD THEORY,''
Nucl.\ Phys.\  B {\bf 345}, 551 (1990).

\bibitem{zwiebach}
B.~Zwiebach,
``Closed string field theory: Quantum action and the B-V master equation,''
Nucl.\ Phys.\ B {\bf 390}, 33 (1993)
[arXiv:hep-th/9206084].

\bibitem{polchinski}
J.~Polchinski,
``String theory. Vol. 1: An introduction to the bosonic string,''
{\it  Cambridge, UK: Univ. Pr. (1998) 402 p}




\end{thebibliography}
\end{document}